\documentclass[useAMS,usenatbib]{mn2e}
\usepackage{txfonts}
\usepackage[latin1]{inputenc}
\usepackage[T1]{fontenc}
\usepackage{mathptmx}
\usepackage[scaled]{helvet}
\usepackage{courier}
\usepackage{aecompl} 
\usepackage{longtable}  
\usepackage{rotating}
\usepackage{natbib}
\usepackage{graphicx}
\usepackage{graphics}
\usepackage{psfrag}
\usepackage{amssymb}
\usepackage{multicol}
\usepackage{multirow}
\usepackage{lscape}
\usepackage{color}
\usepackage{soul}
\usepackage{array}
\usepackage{longtable}
\bibpunct{(}{)}{;}{a}{}{,}

\def\pdra{$\varphi$~Dra}
\def\pdraAab{$\varphi$~Dra~Aab}

\def\oc{\textit{O$-$C}}
\usepackage{amsmath}

\title[Analysis of the multiple system $\boldsymbol\varphi$~Draconis]{Analysis of the multiple system with CP component $\boldsymbol\varphi$~Draconis}

\author[J. Li\v{s}ka]{J. Li\v{s}ka\thanks{jiriliska@post.cz}
			   \\
Department of Theoretical Physics and Astrophysics, Masaryk University,
Kotl\'a\v{r}sk\'a 2, 611\,37 Brno, Czech Republic}

\begin{document}

\date{Received ..., 2015; }

\pagerange{\pageref{firstpage}--\pageref{lastpage}} \pubyear{2016}

\maketitle

\label{firstpage}

\begin{abstract}
{The star \pdra~comprises a spectroscopic binary and a third star that together form a visual triple system. It is one of the brightest chemically peculiar (CP) stars of the upper main sequence. Despite these facts, no comprehensive study of its multiplicity has been performed yet. In this work, we present a detailed analysis of the triple system based on available measurements. We use radial velocities taken from four sources in the literature in a re-analysis of the inner spectroscopic binary (Aab). An incorrect value of the orbital period of the inner system Aab about 27 days was accepted in literature more than forty years. A new solution of orbit with the 128-day period was determined. Relative position measurements of the outer visual binary system (AB) from Washington Double Star Catalog were compared with known orbital models. Furthermore, it was shown that astrometric motion in system AB is well described by the model of Andrade (2005) with a 308-year orbital period. Parameters of A and B components were utilized to estimate individual brightness for all components and their masses from evolutionary tracks. Although we found several facts which support the gravitational bond between them, unbound solution cannot be fully excluded yet.}

\end{abstract}
  

\section{Introduction}
Chemically peculiar (CP) stars play a key role as a test of evolutionary models due to their anomalous atmospheric composition. Many of them are periodic variable stars with periods corresponding to rotational periods \citep{adelman2002}. Therefore, CP stars are very helpful for a study of rotational periods and their stability among main-sequence stars. Preferably, stellar parameters would be known with high precision for such studies, especially  the stellar mass. Mass of the star is one of the most important parameters for establishing of the evolutionary status and can be determined with high accuracy when the star belongs to a binary system. Information about binarity is also necessary for a study of brightness variations (even if it is not an eclipsing system). An additional light source in the system influences the observed amplitude of variability (decrease in observed amplitude) and observed spectra (depths and shapes of spectral lines, appearance of additional lines). Another component also affects radial velocities (hereafter RVs) of the studied star.  
	
Our target \pdra~(HD~170000 = HIP~89908 = HR~6920) is one of the brightest CP stars ($V =4.22$\,mag). For a long time, it has been known as a variable star \citep{winzer1974,schoeneich1976,kukarkin1977} with relatively low amplitude in optical bands (0.02\,--\,0.03\,mag) and rotational period about 1.7\,d e.g. 1.7164 \citep{winzer1974}, 1.71646(6)\,d \citep{musielok1980}. We selected this star for calculation of model of its atmosphere, which was subsequently used for light curve reconstruction \citep{prvak2014,prvak2015}. The availability of abundance maps from Doppler imaging in \citet{kuschnig1998}, which were used for calculation of atmospheric models, was the main criterion for the selection of this star. The opportunity to compare our light curve models with high-accurate 10-colour photometry obtained by \citet{musielok1980} belongs to the great advantages. The preliminary results from this effort, in which the main chemical elements known from spectra are included in the model of the synthetic atmosphere, are very promising \citep{prvak2014}. A similar approach was used e.g. for  star HD~37776 \citep{krticka2007} and HR~7224 \citep{krticka2009}.

Nevertheless, \pdra~is a multiple stellar system and several basic parameters, such as the period of the inner binary system, are highly uncertain (see below). In this paper, we performed a detailed analysis of the multiple system to adequately estimate the influence of the multiplicity on our light curve modelling, which is presented in \citet{prvak2015}.  

The object \pdra~is probably a triple system with components Aa, Ab and B \citep{tokovinin2008} but a fourth component C (optical) is also present in the Washington Double Star Catalog \citep[WDS,][]{mason2001}. Outer binary system AB creates a visual binary with known brightness for both components from Hipparcos satellite $H_{\rm p, A} = 4.455(3)$\,mag and $H_{\rm p, B} = 5.900(10)$\,mag \citep{esa1997}. Their orbit with very long period was calculated several times and the last published orbital parameters from \citet{andrade2005} include orbital period $P_{\rm AB} = 307.8$ years, angular projection of semi-major axis $a_{\rm AB} = 0.752\,''$, and high eccentricity $e_{\rm AB}=0.752$. Main A component is an SB1-type spectroscopic binary (Aa, Ab) with contradictory orbital period of 26.768(7) days \citep{abt1973} or 127.85 days \citep{beardsley1969}. 

The star \pdra~is a frequently studied, but no comprehensive study of the multiple system exists.  Furthermore, several catalogues adopted the orbital period of \citet{abt1973} despite the fact that it was determined only from 10 measurements and the authors themselves doubted its correctness because this period is not compatible with any of older datasets.

We re-analysed the available RV measurements to obtain correct orbital parameters of the inner binary system Aab (Sect.~\ref{SpecBin}). Thereafter, we used relative position measurements of B component from WDS and verified results from \citet{andrade2005}, Sect.~\ref{VisBin}. Evidence for the gravitational bond of the binary is presented in Sect.~\ref{VisBinBound}. In addition, we consider an unbound explanation (Sect.~\ref{VisBinUnbound}). We briefly discuss component C and whether it is bound to components A and B in Sect.~\ref{OpticBin}. We also analyse the whole triple system together (Sect.~\ref{WholeSystem}). Our final results are summarized and discussed in Sect.~\ref{summarysection}.

\section{Inner spectroscopic binary system \pdraAab}\label{SpecBin}

\subsection{Available RV measurements}\label{SpecBinLiterature}

The first spectroscopic observations of \pdra~\citep{maury1897} revealed a peculiar spectrum with lines of hydrogen (Balmer series), He~I (4026.4 \AA), and Si~II (4128.5, 4131.4 \AA). RVs of \pdra~were measured with low-resolution spectrographs (from composed spectra with fainter B component\footnote{B component is fainter about 1.5\,mag than A component (Sect.~\ref{brightness}) and angular distance between both components was under 0.5\,$''$ during 20$^{\rm th}$ century (Sect.~\ref{VisBin}).}) several times during the 20$^{\rm th}$ century \citep[see Table~\ref{tab-rvlist},][]{frost1909,hnatek1914,campbell1928,frost1929,harper1937,beardsley1969,abt1973}. Unfortunately, modern high-quality spectra are almost absent. Exceptions are spectra from \citet{kuschnig1998} used for Doppler imaging (no RV) and one spectrum obtained by \citet{takeda1999} with one RV value. 

\begin{center}
\begin{table}
\caption{List of available RV measurements of \pdra. Used RV measurements are above the central line.}
\centering
\begin{tabular}{lcc}
\hline
Author					& $N_{\rm obs}$	& Year of obs.\\
\hline
\citet{frost1909}$^{*}$		& 4				& 1906 -- 1909\\ 
\citet{frost1929} 			& 28 + 4$^{**}$	& 1920 -- 1922\\
\citet{beardsley1969}  		& 39			& 1911 -- 1914\\
\citet{abt1973} 			& 10			& 1965 -- 1966\\
\hline
\citet{hnatek1914}$^{*}$	& 3				& 1913\\
\citet{campbell1928} 		& 4				& 1896 -- 1905\\
\citet{harper1937} 			& 2				& 1927\\
\citet{takeda1999}			& 1				& 1992\\
\hline
\end{tabular}\label{tab-rvlist}\\
{\bf Notes:} $^{(*)}$ Double lines were detected, $^{(**)}$ four RV values were obtained by re-analysing of spectra from \citet{frost1909}.\\
\end{table}
\end{center}

Variations in RV were detected by \citet{frost1909}, they also found doubling of lines on 3 plates from which they deduced binarity of the A component. The same plates were re-analysed by Kohl \citep[values were published in][]{frost1929}, he measured RVs only from centres of spectral lines. The occurrence of double lines in 2 spectra was also published by \citet{hnatek1914}. However, all photographic double line measurements of \pdra~are probably only artefacts not related to the binarity as discussed \citet{frost1929}. Nevertheless, variations in the shape of lines are known, but only from high-resolution spectroscopy and are caused by CP variability \citep{kuschnig1998}. \citet{frost1929} also obtained new RV measurements with high scatter around mean value $-19.2$\,km\,s$^{-1}$ but they did not mention any explanation.

\citet{beardsley1969} published 39 RV measurements from spectra obtained on Allegheny observatory in 1911\,--\,1914. He confirmed the variability in RVs with semi-amplitude of $30$\,km\,s$^{-1}$ and he explained it as an orbit of unseen component with period of 127.85\,d (SB1-type binary). Beardsley also determined other orbital parameters of this inner binary system Aab from RV modelling (see Table~\ref{tab-rvorbit}). In addition, he mentioned that residuals after subtraction of RV binary model are highly scattered and he attributed it to an unspecified short-period variation. 

\citet{abt1973} obtained 10 RV measurements with higher accuracy on Kitt Peak observatory in 1965\,--\,1966. They established shorter orbital period of 26.768(7) days based only on their values. They noted that their orbital parameters (Table~\ref{tab-rvorbit}) are not usable for extrapolation of older measurements and thus the real period could be different. Despite the fact, their period is frequently adopted in literature \citep{pourbaix2004,tokovinin2008,ducati2011}.

\subsection{Analysis of RV measurements}\label{SpecBinAnalysis}

\begin{figure}
            \centering
            \resizebox{1.0\hsize}{!}{\includegraphics{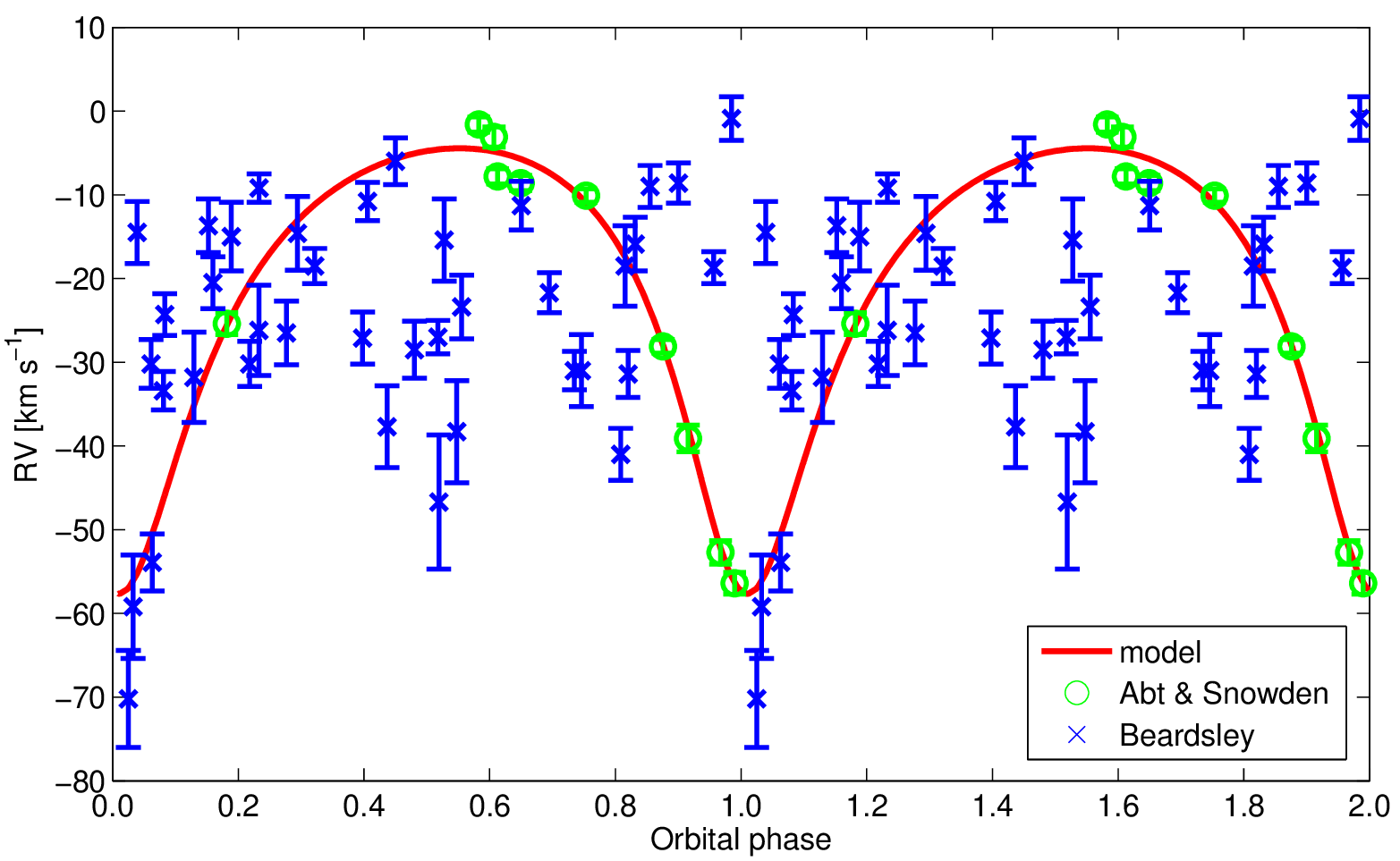}}
            \resizebox{1.0\hsize}{!}{\includegraphics{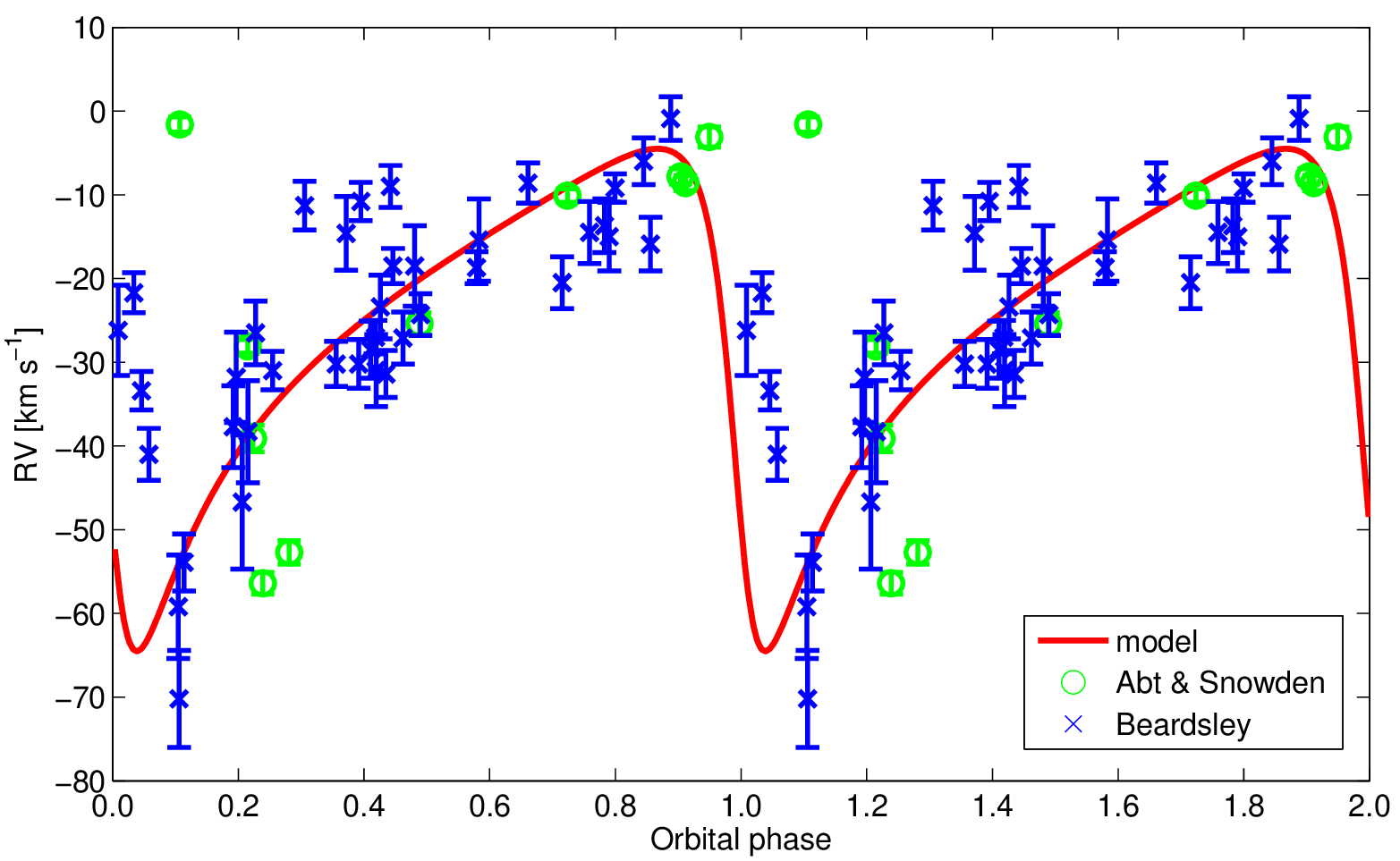}}
            \caption{Observed RV curves for the inner binary system Aab obtained by \citet{abt1973} and \citet{beardsley1969} together with model curves based on orbital parameters from \citet{abt1973} (upper panel) and \citet{beardsley1969} (lower panel).}
            \label{fig:rv_beardsleyabt1}
        \end{figure}

Difference between orbital periods 26.768(7)\,d \citep{abt1973} and 127.85\,d \citep{beardsley1969} is remarkable. It is necessary to determine the correct value. In the beginning, we used datasets from \citet{beardsley1969} and \citet{abt1973} and compared them with the models of RV curves for both sets of orbital elements. Model calculated with elements from \citet{abt1973} is not consistent with data from \citet{beardsley1969}, but fits well their own measurements (Fig.~\ref{fig:rv_beardsleyabt1}, upper panel). On the other hand, model based on parameters from \citet{beardsley1969} describes variations in both datasets much better (Fig.~\ref{fig:rv_beardsleyabt1}, bottom panel), nevertheless, model is slightly shifted in phase for both measured RV curves. It is evident from Fig.~\ref{fig:rv_beardsleyabt1} and also from sum of squares of the residuals between both datasets and models ($R_{\rm Aab} = 10142$ and $16407$ for parameters from \citet{beardsley1969} and \citet{abt1973}, respectively), that period of \citet{beardsley1969} is much closer to the correct value than the \citet{abt1973} one. 

\begin{figure}
            \centering
             \resizebox{1.0\hsize}{!}{\includegraphics{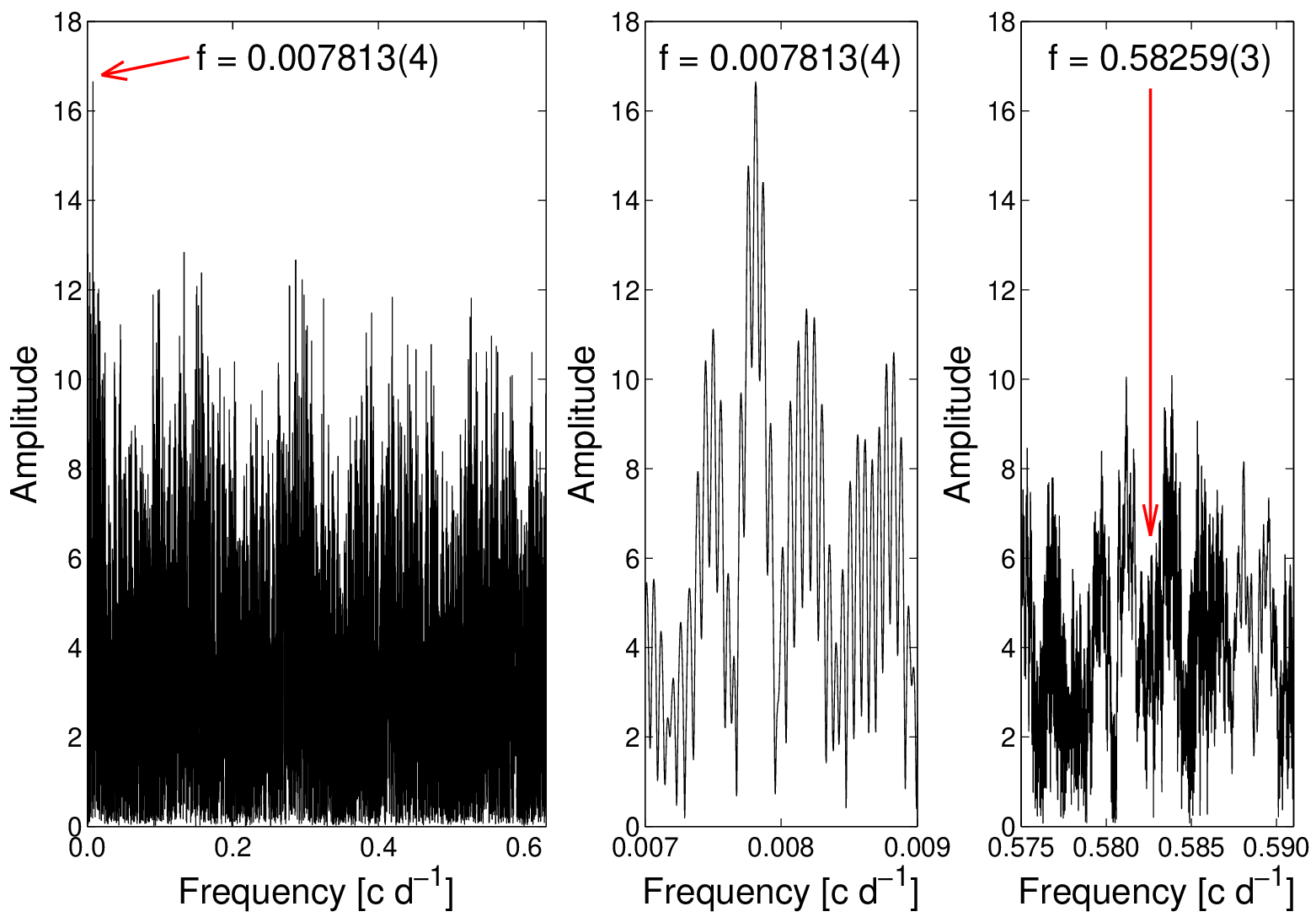}}
            \caption{Frequency spectrum obtained from RV measurements from \citet{frost1929}, \citet{beardsley1969}, and \citet{abt1973} using \textsc{Period04} software shows the strongest frequency 0.007813(4)\,c\,d$^{-1}$ (period 127.99(7)\,d). Full spectrum is displayed in the left panel and detail around the strongest frequency is in the middle panel. Frequency 0.58259(3)\,c\,d$^{-1}$ corresponding to the period of the CP variability 1.71646(6)\,d \citep{musielok1980} is not visible in the spectrum (right panel).}
            \label{fig:rv_frequency}
        \end{figure}

Subsequently, we performed a frequency analysis of RV measurements from the most extensive datasets\footnote{We ignored the low numerous datasets from Table~\ref{tab-rvlist}, because are not statistically significant.} \citep{frost1929, beardsley1969, abt1973} using \textsc{Period04} software \citep{lenz2005}. We found the strongest frequency 0.007813(4)\,c\,d$^{-1}$ (period 127.99(7)\,d) which is very close to Beardsley's value 127.85\,d (see Fig.~\ref{fig:rv_frequency}). We also tested the value of the period of 1.71646(6)\,d, which could indicate a connection with the period of CP variability (rotation of the CP star). This value was not found in these RVs (right panel of Fig.~\ref{fig:rv_frequency}). 

Therefore, we adopted the binary explanation and used non-linear least-squares fitting \citep{mikulasek2007,mikulasek2013,mikulasek2015} for determination of the orbital parameters based on equations from \citet{irwin1952}. Kepler's equation was solved iteratively by Newton's method. All derivatives were calculated analytically. We used an approach similar to that of, e.g., \citet{vanhamme2007}.

Dataset from \citet{frost1929} was published without measurement uncertainties \citep[data contains also the re-analysed RVs from spectra obtained by][]{frost1909}. The other used datasets have underestimated uncertainties, as we found from scatter of residuals after subtraction of the best model (scatter was higher than the mean value of uncertainties). Therefore, we used different uncertainties. Their values were iteratively calculated during the fitting process according to the root-mean-square scatter of residuals between model and each dataset. We found mean RV  uncertainty of 11.9\,km\,s$^{-1}$ for \citet{frost1929}, 7.3\,km\,s$^{-1}$ for \citet{beardsley1969}\footnote{It is 2.04 times higher than mean uncertainty of original RVs.}, and 1.6 km\,s$^{-1}$ for \citet{abt1973}\footnote{It is 1.44 times higher than mean uncertainty of original RVs.}, respectively. 

\begin{figure}
            \centering
            \resizebox{1.0\hsize}{!}{\includegraphics{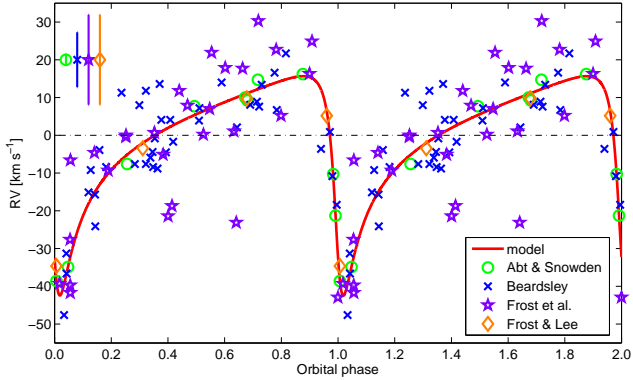}}
            \caption{RV variations for the inner spectroscopic binary system Aab obtained from four datasets (individual $\gamma$-velocities were subtracted) with calculated measurement uncertainties in left-top corner and model curve based on our orbital parameters from Table~\ref{tab-rvorbit}.
            }
            \label{fig:rv_fit}
        \end{figure}

The best found solution with the lowest sum of squares of the residuals $R_{\rm Aab} = 6420.7$ has weighted value $\chi^{2} = 78$ with normalised value $\chi_{\rm R}^{2} = \chi^{2}/(N-g) = 1.08(17)$. The number of measurements $N$ is 81 and the number of fitted parameters $g$ is 9. The final parameters were used to determine their uncertainties by bootstrap-resampling method. Errors given in Table~\ref{tab-rvorbit} correspond to $1\,\sigma$. The observed variations in RVs are well described by our model (Fig.~\ref{fig:rv_fit}), which is in good agreement with $\chi_{\rm R}^{2} \sim 1$ (distribution of residual values is similar to normal distribution).

Our modelling of the system Aab orbit allowed to find more accurate value of orbital period $P_{\rm Aab\,orbit} = 127.9902^{+37}_{-30}$\,d with high eccentricity $e_{\rm Aab} \sim 0.7$. Observed variation in RV can be described by semi-amplitude of RV $K_{\rm Aa} \sim 29$\,km\,s$^{-1}$, and $\gamma$-velocities determined for each datasets\footnote{RVs from \citet{frost1929} were divided into two groups according to the time of observations -- old values from \citet{frost1909} and later ones -- to study variations in $\gamma$-velocity. Their uncertainties were adopted the same.}. In addition we established the mass function $f(m_{\rm Ab}) \sim 0.12$\,M$_{\odot}$ and the lowest mass of Ab component $M_{\rm Ab\,min} \sim 1.36$\,M$_{\odot}$. This mass limit was computed for inclination angle of $90\,^{\circ}$ and adopted mass of CP star $M_{\rm Aa} = 3.3$\,M$_{\odot}$ according to values 3.29(17)\,M$_{\odot}$ \citep{allende1999} and 3.30(8)\,M$_{\odot}$ \citep{kochukhov2006}. The projection of semi-major axis of the Aa component is $a_{\rm Aa}\sin i \sim 0.24$\,au. The orbital parameters allowed us to predict semi-amplitude of Light Time Effect (LiTE) $A_{\rm LiTE,Aa} \sim 0.0012$\,day which influences rotational period of CP component (see Sect.~\ref{mass}). Our determined orbital parameters are quite similar as values found by \citet{beardsley1969}.

\begin{center}
\begin{table}
\caption{Determined parameters for the inner spectroscopic binary pair \pdra~Aab.}
\begin{center}
\tabcolsep=3.5pt
\begin{tabular}{lcc|c}
\hline
Author 									& Beardsley 	& Abt\,\&\,Snowden 	& This paper\\
Pub. year       						& (1969)		& (1973) 			& -- \\
\hline
$P_{\rm Aab\,orbit}$ [d] 				& 127.85 		& 26.768(7) 		& 127.9902$^{+37}_{-30}$\\
$T_{0\,\rm Aab}$ [HJD]					& 2413823.7		& 2438853.6(1.6)	& 2413825.79$^{+76}_{-91}$\\ 
$e_{\rm Aab}$							& 0.60 			& 0.39(4) 			& 0.707$^{+21}_{-15}$\\  
$\omega_{\rm Aab} [^{\circ}]$ 			& 120 			& 171(17) 			& 130.8$^{+6.0}_{-5.5}$\\ 
$K_{\rm Aa}$ [km\,s$^{-1}$] 			& 30			& 26.6(2.1)			& 29.0$^{+1.3}_{-1.2}$\\
$\gamma_{\rm Be}$ [km\,s$^{-1}$] 		& $-25.5$		& --				& $-22.6^{+1.1}_{-1.3}$\\
$\gamma_{\rm AbSn}$ [km\,s$^{-1}$] 		& --			& $-20.8(8)$		& $-17.8^{+1.1}_{-0.9}$\\
$\gamma_{\rm FrLe}$ [km\,s$^{-1}$] 		& --			& --				& $-16.0^{+1.7}_{-1.6}$\\
$\gamma_{\rm FrBaSt}$ [km\,s$^{-1}$] 	& --			& --				& $-15.9^{+2.3}_{-2.4}$\\
$R_{\rm Aab}$							& --			& --				& 6420.7\\
$\chi^2$ 								& --			& --				& 78\\
$\chi_{\rm R}^2$ 						& --			& --				& 1.08(17)\\
$N$ 									& 39 			& 10 				& 81\\
\hline
$a_{\rm Aa}\sin i$ [au] 				& 0.282$^{*}$	& 0.0603			& 0.2421$^{+91}_{-87}$\\ 
$A_{\rm LiTE,Aab}$ [d] 					& 0.00155$^{*}$	& 0.00032$^{*}$		& 0.001241$^{+73}_{-79}$\\ 
$f(m_{\rm Ab})$ [M$_{\odot}$] 			& 0.183$^{*}$ 	& 0.0408			& 0.116$^{+13}_{-13}$\\ 
$M_{\rm Ab\,min}$ [M$_{\odot}$]		 	& 1.65$^{*}$	& 0.90$^{*}$ 		& 1.359$^{+63}_{-61}$\\
\hline
\end{tabular}\label{tab-rvorbit}\\
\end{center}
{\bf Notes:} $^{(*)}$ -- parameters were calculated by us using values from original study.\\
{\bf Parameters:} $P_{\rm Aab\,orbit}$ -- orbital period, $T_{0\,\rm Aab}$ -- time of periastron passage, $e_{\rm Aab}$ -- numerical eccentricity, $\omega_{\rm Aab}$ -- argument of periastron, $K_{\rm Aa}$ -- semi-amplitude of RV variation, $\gamma$ -- systemic RV of mass-centre of binary, lower indices mark used source of observations -- Be \citep{beardsley1969}, AbSn \citep{abt1973}, FrLe \citep{frost1909}, and FrBaSt \citep{frost1929}, $R_{\rm Aab}$ -- sum of squares of the residuals (non-weighted), $\chi^2$ -- quality indicator of the fit (value of $R_{\rm Aab}$ normalised by uncertainties), $\chi_{\rm R}^{2}$ -- value of $\chi^{2}$ normalised by numbers of RVs and free parameters, $N$ -- number of used RV measurements, $a_{\rm Aa}\sin i$ -- projection of semi-major axis of the Aa component, $A_{\rm LiTE,Aa}$ -- semi-amplitude of variations in $O\!-\!C$ diagram cause by LiTE, $f(m_{\rm Ab})$ -- mass function, $M_{\rm Ab\,min}$ -- the lowest mass of Ab component (inclination $90\,^{\circ}$, mass $M_{\rm Aa} = 3.3$\,M$_{\odot}$). 
\end{table}
\end{center}

\section{Outer visual binary system \pdra~AB }\label{VisBin}

\subsection{Analysis of available data}\label{VisBinLit}
The visual binary system \pdra~AB also named WDS~J18208+7120AB = CCDM J18208+7120AB = STT~353AB is known since the 19$^{\rm th}$ century. The first position measurement of B relative to A component was performed in 1843 \citep[WDS,][]{mason2001}. Up to now, under 100 averaged relative measurements of position for the B component exist. The orbit of the visual binary AB was computed several times \citep{olevic1975, olevic1990, andrade2005} and parameters from these studies are summarized  for comparison in Table~\ref{tab-visualorb}. We added values of the absolute semi-major axis of the orbit $a_{\rm AB}$ which were calculated from their angular projections and the parallax of \pdra~adopted from Hipparcos measurements $\pi = 10.77(38)$\,mas \citep{vanleeuwen2007}. It allows us to determine the mass of the whole visual binary system $M_{\rm AB}$.

\begin{center}
\begin{table}
\begin{center}
\caption{Determined parameters for the outer visual binary system \pdra~AB.}
\begin{tabular}{lcccc}
\hline
Author 								& Olevic$^{*}$	& Olevic \& Catovic & Andrade	\\
Pub. year       					& (1975)	 	& (1990) 			& (2005)	\\
\hline
$P_{\rm AB\,orbit}$ [yr] 			& 271.7			& 401.85			& 307.8   	\\
$T_{0 \,\rm AB}$ [yr]				& 1720.750		& 1767.69			& 2116.6  	\\
$e_{\rm AB}$						& 0.440 		& 0.236				& 0.752   	\\
$\Omega_{\rm AB}$ $[^{\circ}]$ 		& 72.25 		& 66.9				& 70.3    	\\
$a_{\rm AB}$ $['']$ 				& 0.392			& 0.614				& 0.965   	\\
$i_{\rm AB}$ $[^{\circ}]$ 			& 118.96 		& 103.6				& 95.6    	\\
$\omega_{\rm AB}$ $[^{\circ}]$ 		& 201.25 		& 282.7				& 275.0   	\\
Last Obs. [yr] 						& 1968 			& 1986.405			& 1998.679	\\
\hline
$a_{\rm AB}$ [au]$^{**}$			& 36.40			& 57.01				& 89.60	  	\\
$M_{\rm AB}$ [M$_{\odot}$]$^{**}$	& 0.65			& 1.15				& 7.59	  	\\
$R_{\rm AB,92}$\!$^{**}$ 			& 15.51			& 5.90				& 5.04\\
\hline
\end{tabular}\label{tab-visualorb}
\end{center}
{\bf Notes:} $^{(*)}$ -- parameters were used from the Fourth Catalog of Orbits of Visual Binary Stars \citep{worley1983}, 
$^{(**)}$ -- parameters were calculated by us using values from original study.\\
{\bf Parameters:} $P_{\rm AB\,orbit}$ -- orbital period, $T_{0 \,\rm AB}$ -- time of periastron passage, $e_{\rm AB}$ -- numerical eccentricity, $\Omega_{\rm AB}$ -- position angle of node, $a_{\rm AB}$ -- angular value of total semi-major axis, $i_{\rm AB}$ -- inclination angle of the orbit, $\omega_{\rm AB}$ -- argument of periastron, Last. Obs. -- date of the last position measurements, $a_{\rm AB}$ -- absolute value of total semi-major axis -- parallax was adopted $\pi = 10.77(38)$\,mas \citep{vanleeuwen2007}, $M_{\rm AB}$ -- total mass of AB system, $R_{\rm AB,92}$ -- sum of squares of the residuals from model of relative orbit and 92 available positions.
\end{table}
\end{center}

The orbital parameters given in the literature differ significantly (Table~\ref{tab-visualorb}), and so do the calculated parameters $a_{\rm AB}$ and $M_{\rm AB}$. We have not found any comparison of the consistency of the three orbital models with the observational data. Furthermore, the last model of the orbit was published in 2005, and the last observation used in the mentioned work was obtained in 1998. We decided to verify all three models to select the model which successfully describes all available data. 

\begin{figure}
            \centering
            \resizebox{1.0\hsize}{!}{\includegraphics{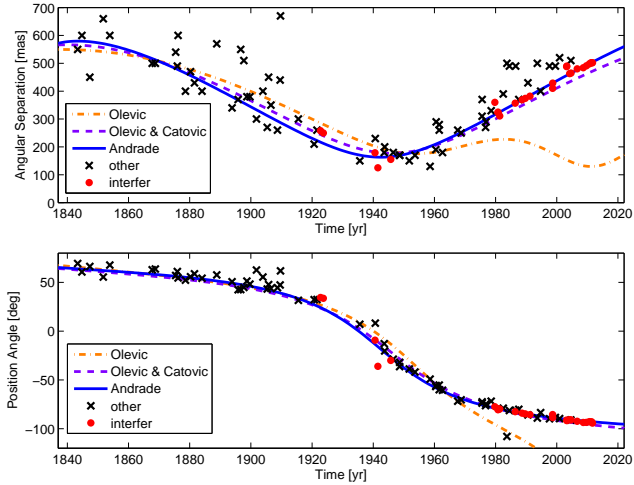}}
            \caption{Measurements of angular separation and position angle of B component relative to A component adopted from WDS together with model curves based on orbital parameters from \citet{olevic1975}, \citet{olevic1990} and \citet{andrade2005} given in Table~\ref{tab-visualorb}.
           }
            \label{fig:position_measur}
        \end{figure}

\begin{figure}
            \centering
            \resizebox{1.0\hsize}{!}{\includegraphics{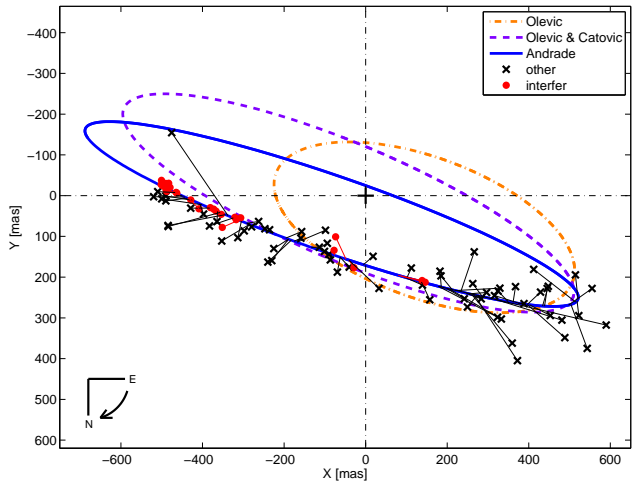}}
            \caption{Relative orbit of the outer visual binary system \pdra~AB reconstructed from relative position measurements listed in WDS and model curves based on orbital parameters from \citet{olevic1975}, \citet{olevic1990} and \citet{andrade2005} given in Table~\ref{tab-visualorb}. 
           }
            \label{fig:visual_orbit}
        \end{figure}

We used data from WDS (version from 2014, April, 30); 92 pairs of relative position parameters (angular separation $\varrho$ and position angle $\theta$) were obtained in the time interval from 1843.38 to 2011.69. The three values of position angle from \citet{comstock1896} are probably shifted by 180\,$^{\circ}$ (the corrected values were used in our calculation). We displayed the time dependences of the parameters $\varrho$, $\theta$ together with known models (Fig.~\ref{fig:position_measur}) and reconstructed the relative orbit (Fig.~\ref{fig:visual_orbit}). Interferometric values, which should have the highest accuracy from all measurements, are listed in the Fourth Catalog of Interferometric Measurements of Binary Stars\footnote{http://www.usno.navy.mil/USNO/astrometry/optical-IR-prod/wds/int4} \citep[INT4,][]{hartkopf2001} and are marked in figures as ``interfer''. With respect to these figures, models based on parameters from \citet{olevic1975} and \citet{olevic1990}\footnote{Model from \citet{olevic1990} does not describe the latest variations in $\varrho(t)$ and relative orbit (Fig.~\ref{fig:visual_orbit}).} can be ruled out because models disagree with the latest observations. Calculated sum of squares of the residuals\footnote{Non-weighted, the uncertainties of most measurements were not known.} $R_{\rm AB,92}$ for models of relative orbits and all available measurements (Table~\ref{tab-visualorb}) are also worse than for the latest study \citep{andrade2005}. Models based on parameters from \citet{andrade2005} well describe all dependencies and we adopted his parameters for our subsequent analysis. Despite the fact, these parameters are still very uncertain because the whole orbit was not observed yet due to the long period of about 300 years. Moreover, unbound explanation discussed below is also possible.

\subsection{Binary System \pdra~AB -- evidence for a gravitationally bound system}\label{VisBinBound}

The shape of changes in the relative position between A and B component (Fig.~\ref{fig:visual_orbit}) is ``almost rectilinear relative motion'' (WDS). Therefore, we propose that the system AB might not be bound and the fainter star only passes around the \pdra. Our theory is supported through the high eccentricity $e_{\rm AB} = 0.752$ found by \citet{andrade2005} and fast-growing angular and absolute semi-major axis $a_{\rm AB}$ in different studies (36.40\,au in 1975, 57.01\,au in 1990, 89.60\,au in 2005, see Table~\ref{tab-visualorb})\footnote{In case of unbound system, the eccentricity and semi-major axis can grow continually with longer time base to describe observed changes.}. Therefore, we were looking for other evidence that could confirm or disprove the gravitational binding.

Proper motions were mostly measured for the whole system \pdra~AB (dominant A component) \citep[e.g.,][]{roeser2010,zacharias2012,zacharias2013}. Probably only Hipparcos observations have yielded independent values for the A and B components \citep{esa1997}. The similarity of the values supports the hypothesis that the two components are bound (see Table~\ref{tab-propermotion}).

The parallaxes for both components in catalogue Double and Multiples: Component solutions \citep{esa1997} are the same $\pi=11.28(48)$\,mas and were measured only for the brighter A component. The corrected value of parallax $\pi = 10.77(38)$\,mas \citep{vanleeuwen2007} is given only for the whole system. Therefore, the both components may or may not be in the same distance from the Sun.

The position of A and B components in HR diagram is promising. When we estimate that both stars have roughly the same age, the observed brightness difference between them is caused only through different luminosity. Therefore, the stars should lie on the one isochrone. We found that positions of A and B components in HR diagram (for the same distance from Sun) correspond to different isochrones, but the distance between the lines is small (details in Sect.~\ref{HRdiagram}).

We did not find in literature any information about spectroscopic measurements of the B component or its RV.

\begin{center}
\begin{table}
\centering
\tabcolsep=5pt
\caption{Proper motions for the whole system \pdra~A+B and for individual components.}
\begin{tabular}{cccc}
\hline
Comp.					& pmRA				& pmDEC				& Author\\
\pdra~					& [mas\,yr$^{-1}$]	& [mas\,yr$^{-1}$]	& Pub. year\\
\hline
A+B						& $-9.6(2.0)$		& $34.6(2.2)$		& \citet{roeser2010}\\ 
A+B						& $-5.9(1.0)$		& $35.8(1.0)$		& \citet{zacharias2012,zacharias2013}\\ 
A						& $-5.91(0.54)$		& $35.79(0.57)$		& \citet{esa1997}\\ 
B						& $-13.09(1.24)$	& $33.76(1.67)$		& \citet{esa1997}\\ 
C						& $-4.2(5.2)$		& $-7.6(5.2)$		& \citet{roeser2010}\\ 
C						& $-5.6(2.8)$		& $-6.0(1.9)$		& \citet{zacharias2012,zacharias2013}\\ 
\hline
\end{tabular}\label{tab-propermotion}
\end{table}
\end{center}

\subsection{Binary system \pdra~AB -- test of gravitationally unbound solution}\label{VisBinUnbound}

\begin{figure}
            \centering
            \resizebox{1.0\hsize}{!}{\includegraphics{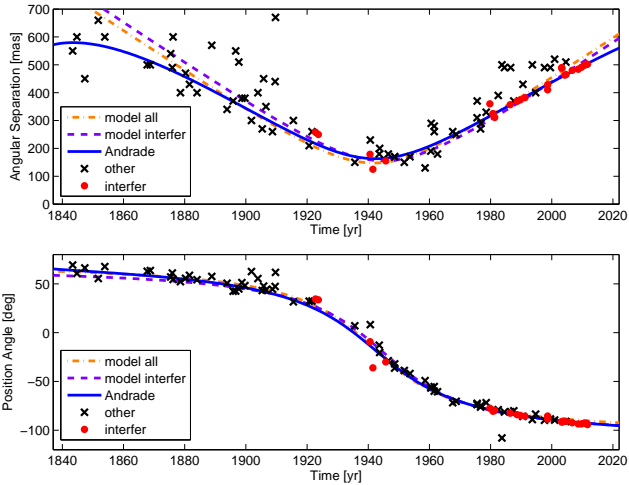}}
            \caption{Measurements of angular separation and position angle of B component relative to A component adopted from WDS. Graph contains model curves based on unbound scenario for all data and only for interferometric measurements (parameters given in Table~\ref{tab-visualunbound}), and models based on orbital parameters from \citet{andrade2005}.
           }
            \label{fig:position_measur_lin}
        \end{figure}

\begin{figure}
            \centering
            \resizebox{1.0\hsize}{!}{\includegraphics{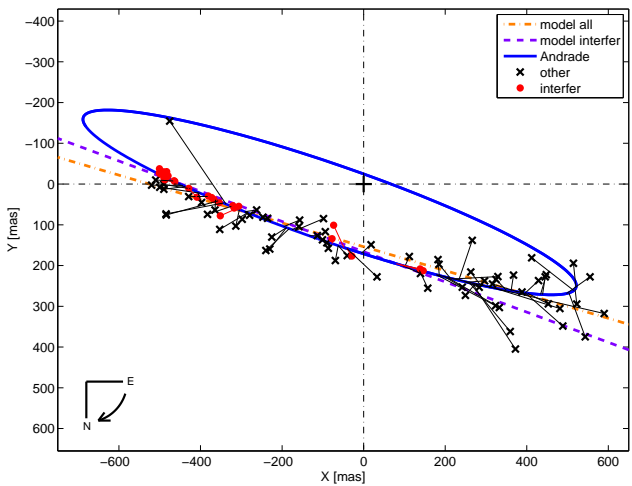}}
            \caption{Relative orbit of the visual binary system \pdra~AB reconstructed from relative position measurements in WDS. Graph contains model curve based on orbital parameters from \citet{andrade2005}, and models of unbound scenario for all data and only for interferometric measurements (parameters given in Table~\ref{tab-visualunbound}). The lines describe situation when stars A and B only accidentally meet in sky and not in the space. 
           }
            \label{fig:visual_orbit_lin}
        \end{figure}

As was mentioned, the shape of the visual relative motion of the A and B stars indicates the possibility that the two components are not bound. We applied the least-squares method (only non-weighted, because uncertainties for most of measurements are not known) to find the model of the relative proper motion between both stars which could explain the observed changes in position. We used values of $\varrho(t)$ and $\theta(t)$ from WDS and we fitted values of velocity B star relative to A star in both directions $v_{\rm x,B\rightarrow A}$, $v_{\rm y,B\rightarrow A}$, time of minimal distance between both stars $t_{0}$, and corresponding position B star relative to A star $x_{\rm B\rightarrow A}(t_{0})$, $y_{\rm B\rightarrow A}(t_{0})$. Mentioned parameters were determined on the whole dataset and on the high-accurate measurements (all interferometric measurements from INT4). The best found model for all measurements (parameters are in Table~\ref{tab-visualunbound}) well describe the observed variations in angular separation and position angle (Fig.~\ref{fig:position_measur_lin}) as same as relative motion (Fig.~\ref{fig:visual_orbit_lin}). Although, we cannot fully exclude gravitationally unbound explanation, model of binary orbit from \citet{andrade2005} describes variations better (sum of squares of the residuals $R_{\rm AB,92}$ = 5.04) than does the unbound model ($R_{\rm B\rightarrow A}$ = 5.72). It is visible, especially in the latest interferometric observations. Therefore, we accepted bound solution as correct.

\begin{center}
\begin{table}
\centering
\tabcolsep=3.5pt
\caption{Our determined parameters for unbound model of the binary pair \pdra~AB.}
\begin{tabular}{lcc}
\hline
Data set										& All measurements 		& Interferometric\\
\hline
$t_{0}$ [yr]									& 1941.8(1.3)			& 1945(2)\\ 
$x_{\rm B\rightarrow A}(t_{0})$ [mas]			& 142(4) 				& 146(5)\\  
$y_{\rm B\rightarrow A}(t_{0})$ [mas]			& $-42(3)$				& $-54(5)$\\ 
$v_{\rm x,B\rightarrow A}$ [mas\,yr$^{-1}$]		& $-2.1(2)$				& $-2.6(2)$\\  
$v_{\rm y,B\rightarrow A}$ [mas\,yr$^{-1}$]		& $-7.1(2)$				& $-7.0(3)$\\ 
Interval [yr] 									& 1843.38 -- 2011.69	& 1922.746 -- 2011.69\\
$N_{\rm obs}$  									& 92					& 26\\
$R_{\rm B\rightarrow A}$ 						& 5.72					& 0.55\\
\hline
\end{tabular}\label{tab-visualunbound}
\end{table}
\end{center}

        \begin{figure}
            \centering \resizebox{0.95\hsize}{!}{\includegraphics{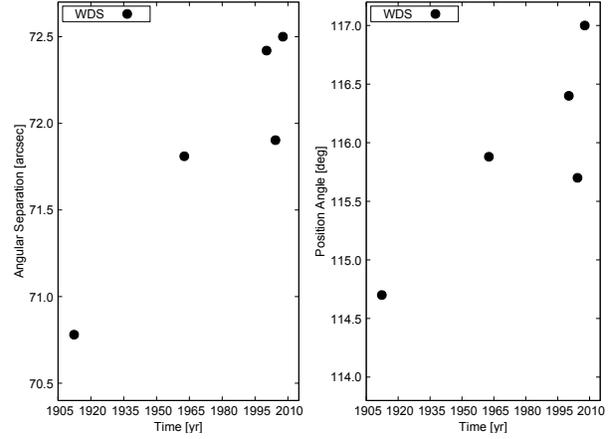}}
            \caption{Measurements of angular separation and position angle of C component relative to AB component adopted from WDS.}
            \label{fig-compC}
        \end{figure}

\section{Optical binary system \pdra~ABC}\label{OpticBin}
WDS \citep{mason2001} contains also the third visual component in \pdra, WDS J18208+7120C = CCDM J18208+7120C = STT 353AB,C = 2MASS J18205889+7119437. This C component with visual brightness about 12.70\,mag ($V$) has only 5 pairs of relative position measurements from 1912.32 to 2007.72 in the used version of WDS. Its position is changed only slightly as seen in Fig.~\ref{fig-compC} and on values of angular separation C from AB ($70.78\,'',\,72.5\,''$) and position angle ($114.7\,^{\circ},\,117\,^{\circ}$) for the first and the last observation given in WDS. 

We did not find the distance of the C component from the Sun in literature. If we adopted the same parallax as for \pdra,~$\pi = 10.77(38)$\,mas \citep{vanleeuwen2007} and circular orbit, then actual minimal distance from system AB is about 6572\,au (for angular separation $70.78\,''$). When we assume the total mass of AB component 7.59\,M$_{\odot}$ and neglecting mass of C component, the orbital period calculated using Kepler's third law is extremely long of 193000\,yr. We also estimate the orbital period from angular changes ($2\,^{\circ}$ per 95.4 years), which give us 17100 years (one order lower value) for completing one orbit. These roughly-estimated values almost exclude gravitational bond between AB system and C component. This opinion is supported via proper motions. The values for C star differ significantly in declination compare to system AB (see Sect.~\ref{VisBinBound}, Table~\ref{tab-propermotion}).

The field around \pdra~belongs among sparse areas of Galaxy ($l  = 101.876\,^{\circ}$, $b = +28.036\,^{\circ}$) with a quite low number of stars. We tried to estimate probability that a star similar as C component (12.7\,mag in $V$) accidentally occurs in area with radius 71\,$''$, therefore total numbers of stars with the same or higher brightness in this area with radius 20, 30, 40, 50, and 60\,$'$ were computed and compared. We found approximately 6\% probability for it using the SIMBAD astronomical database \citep{wenger2000} and given $V$ magnitudes, but our previous experiences with SIMBAD show that this sample is not probably complete. We obtained more realistic estimation from the USNO-B1.0 catalog \citep[][missing $V$-band]{monet2003}, where probability for the same minimal brightness is two times ($R1$-band) or even three times higher ($R2$-band)\footnote{For B component, the probability of random occurrence is almost equal zero -- no other star brighter than 6\,mag up to 60\,$'$ is listed in both sources.}.

The star C is listed as an optical (not bound) component of \pdra~in WDS and our findings also weakly support the hypothesis. In any case, its influence on the main triple system is negligible for our study (low flux contributions in optical band, extremely long-term changes in RV) and we ignore it during other analysis. Our scheme of the multiple system \pdra~known also as ``mobile diagram'' is in Fig.~\ref{fig-hierarchy}.


\section{Analysis of the whole system}\label{WholeSystem}
\subsection{Mass of components}\label{mass}
At first we determined masses of the individual stars. If we adopted the bound visual system AB and parallax $\pi = 10.77(38)$\,mas \citep{vanleeuwen2007}, then we obtain the total mass of the triple system $M_{\rm AB} = 7.59$\,M$_{\odot}$ from visual orbit and parameters determined by \citet{andrade2005}. Mass of the brightest component was adopted $M_{\rm Aa} = 3.3$\,M$_{\odot}$ (see Sect.~\ref{SpecBinAnalysis}). The rest of mass is divided into components Ab and B, unfortunately we can say nothing about mass of B component. However, we have determined lower mass-limit for Ab star $M_{\rm Ab\,min} \sim 1.36$\,M$_{\odot}$ from RV fitting of the inner pair. The exact value of Ab mass depends on the inclination of the orbit Aab. Inclination is not known but may be lower than $90\,^{\circ}$ because the system is not known as an eclipsing one, nevertheless only Hipparcos dataset (106 measurements) covers the whole 128-day period and does not confirm the eclipses. The B component has maximal mass $M_{\rm B\,max} \sim 2.93$\,M$_{\odot}$ from mentioned reasons. 

From this, we predict about three-centuries long cyclic variations in systemic RV of A component caused by B component with maximal semi-amplitude $K_{A} \sim 5.0$\,km\,s$^{-1}$ (for $M_{\rm B\,max}$). Individual values of $\gamma$-velocities for the inner system determined from different RV datasets (Table~\ref{tab-rvorbit}) undergo random time-independent changes and are probably related to systematic errors \citep[especially the maximum deviated value from][]{beardsley1969} and do not correspond expected long-term behaviour. In addition, the mean uncertainties of individual measurements are comparable with this scatter. Orbit of B component also influences observation of periodic effects, such as orbit of the inner binary system or CP variation, through the LiTE. It should be detectable in \oc~diagram with maximal semi-amplitude $A_{\rm LiTE\,A} \sim 0.2$\,days. It was tested by analysis of CP variability, unfortunately only three available photometric datasets covering forty-year long interval are insufficient to show significant variations in \oc~\citep{prvak2015}. The LiTE caused by Aab component ($P_{\rm Aab} \sim 128$\,d, $A_{\rm LiTE,Aa} \sim 0.0012$\,d) is also not apparent in $O\!-\!C$ diagram.

Evolutionary models of stars allow to find mass of the components based on photometry. Construction of HR diagram for visual pair A and B (see Sect.~\ref{HRdiagram}) and best found isochrones enable to estimate mass $M_{\rm Aa} \sim 3.28$\,M$_{\odot}$ and $M_{\rm B} \sim 2.40$\,M$_{\odot}$ for Aa and B component, respectively. In addition, accepting the total mass of the system $M_{\rm AB} = 7.59$\,M$_{\odot}$ from the visual orbit, we obtain mass of Ab component $M_{\rm Ab} \sim 1.91$\,M$_{\odot}$, as the rest of mass. Using mass function known from spectroscopic orbit we calculated inclination angle for Aab system $i_{\rm Aab} \sim 50\,^{\circ}$.

        \begin{figure}
            \centering \resizebox{0.99\hsize}{!}{\includegraphics{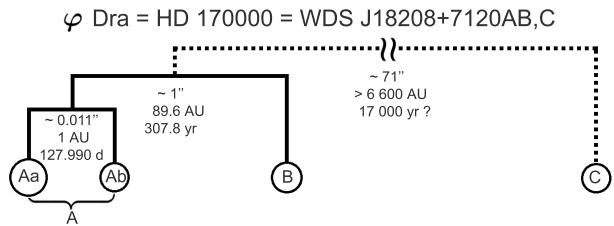}}
            \caption{Mobile diagram of the multiple stellar system \pdra.}
            \label{fig-hierarchy}
        \end{figure}

\subsection{Brightness and spectral types of components}\label{brightness}
Determination of brightness for the individual components and their contribution to the total flux of \pdra~is very difficult due to the lack of precise measurements. The vast majority of the photometric measurements in different passbands \citep[e.g., in][]{musielok1980} correspond to the total brightness of the whole triple system.

The magnitude difference between the components A and B varies from 1.0\,mag to 2.1\,mag according to available literature (see Table~\ref{tab-brightness})\footnote{Most of them are only visual estimated values; angular distance of visual components was under $0.5\,''$ in 20$^{\rm th}$ century and under limit of angular resolution of most classical photometers.}. The values for A and B components from the Hipparcos $H_{\rm p,A} = 4.455(3)$\,mag, $H_{\rm p,B} = 5.900(10)$\,mag \citep{esa1997} and difference between both stars $\Delta H_{\rm p, AB} = H_{\rm p, B} - H_{\rm p, A} = 1.445(10)$\,mag, bring more precise information about A and B components. Data from Hipparcos and Tycho instrument was also used for two-colour photometry of visual components by \citet{fabricius2000}. They found $B_{\rm T,A} = 4.39(1)$\,mag, $B_{\rm T,B} = 5.89(1)$\,mag and $V_{\rm T,A} = 4.48(1)$\,mag, $V_{\rm T,B} = 5.90(1)$\,mag that give us $\Delta B_{\rm T, AB}$ = 1.50(1)\,mag and $\Delta V_{\rm T,AB}$ = 1.42(1)\,mag. We estimate that the B component is probably redder than the A component from the decreasing relative brightness in direction to the longer wavelengths. We found only one other work with original precise measurements of relative brightness in red and infrared passbands $\Delta R_{\rm AB} = 1.42(2)$\,mag, $\Delta I_{\rm AB} = 1.33(3)$\,mag \citep{rutkowski2005}. These values were obtained using speckle interferometry technique and well confirm our assumption. 

\begin{center}
\begin{table}
\caption{Brightness of components \pdra~A and B in different colour bands taken from literature.}
\begin{center}
\def\arraystretch{1.5}
\tabcolsep=1.8pt
\begin{tabular}{lllll}
\hline
$m_{\rm A}$ & $m_{\rm B}$ & $\Delta m_{\rm AB}$ & Passband & Author, Pub. year\\
\hline
4.4			& 6.5		& 2.1$^{*}$			& --			& \citet{campbell1928} \\
4.5			& 6.2		& 1.7$^{*}$			& --			& \citet{moore1932} \\
 --			& --		& 1.0				& vis			& \citet{kuiper1961} \\
 --			& --		& $1.2-2.0$			& vis			& \citet{worley1962} \\
			& 			& avg 1.6(3)		& 				& \\
4.4			& 6.1		& 1.7$^{*}$			& --			& \citet{vanbiesbroeck1974} \\
4.8			& 6.5		& 1.7$^{*}$			& --			& \citet{muller1978} \\
4.40		& 6.10		& 1.70$^{*}$		& $V$ or vis	& \citet{worley1983} \\
4.455(3)	& 5.900(10)	& 1.445(10)$^{*}$	& $H_{\rm p}$	& \citet{esa1997} \\
4.39(1)		& 5.89(1)	& 1.50(1)$^{*}$		& $B_{\rm T}$	& \citet{fabricius2000} \\
4.48(1)		& 5.90(1)	& 1.42(1)$^{*}$		& $V_{\rm T}$	& \citet{fabricius2000} \\
4.2			& 5.7		& 1.5$^{*}$			& $V$ or vis	& \citet{douglass2000} \\
 --			& --		& 1.42(2) 			& $R$			& \citet{rutkowski2005} \\
 --			& --		& 1.33(3) 			& $I$			& \citet{rutkowski2005} \\
\hline
\end{tabular}\label{tab-brightness}
\end{center}
{\bf Notes:} $^{(*)}$ Parameters were calculated by us as difference between $m_{\rm B}$ and $m_{\rm A}$ adopted from the original study.\\
\end{table}
\end{center}

Spectroscopic and most of photometric  measurements of \pdra~provide information about the whole object. Its spectral type was recently estimated as B8V C \citep{gray2014}. According to the colour indices corrected for extinction (see Table~\ref{tab-hrdiagram}) $(B_{\rm T} - V_{\rm T})_{\rm A} = -0.13(1)$\,mag and $(B_{\rm T} - V_{\rm T})_{\rm B} = -0.05(1)$\,mag for A and B component, respectively, we estimated the spectral types B7V for A component and B9V for the B component based on \citet{tsvetkov2008} or Mamajek's list\footnote{http://www.pas.rochester.edu/$\sim$emamajek/\\EEM\_dwarf\_UBVIJHK\_colors\_Teff.txt} (updated July 2015). Determined spectral types from brightnesses are in good accordance with types based on our found masses $M_{\rm Aa} = 3.3$\,M$_{\odot}$ (B8V) and $M_{\rm B\,max} = 2.93$\,M$_{\odot}$ (B9V). The A component is more luminous than B component.

Information about brightness of the Ab component was not found in literature\footnote{High brightness of Ab can influence significantly also Aa itself.}, the presence of Ab is considered only from the RV variations (SB1 binary) and thus, it can be only roughly estimated based on our findings. The CP star Aa is on the main sequence (hereafter MS) with deduced age $2.1\times 10^{8}$ yr \citep{kochukhov2006} or a bit more ($\sim$ 2.45\,--\,3.3$\times 10^{8}$\,yr, Sect.~\ref{HRdiagram}). The other components with lower mass should be also on the MS. Using its mass-limit $M_{\rm Ab\,min} = 1.36$\,M$_{\odot}$ the star would be of spectral type F4V\,--\,F5V or earlier. It should be maximally about 3.5\,mag fainter than Aa component (Mamajek web page).

Masses based on HR diagram bring the same spectral types $M_{\rm Aa} \sim 3.28$\,M$_{\odot}$ (B8V), $M_{\rm B} \sim 2.40$\,M$_{\odot}$ (B9V\,--\,A0V), with difference only for Ab component $M_{\rm Ab} \sim 1.91$\,M$_{\odot}$ (A4V) and its lower difference in brightness $V_{\rm Aa}-V_{\rm Ab} = 2.0$\,mag. Instruments with high angular resolution (angular projection of $a_{\rm Aa}\sin i$ corresponding to 2.6\,mas\footnote{Maximal distance between Aa and Ab components should be about 11\,mas.}) can separate the Ab component from the Aa component and confirm our estimations.

\subsection{HR diagram}\label{HRdiagram}
\begin{figure}
            \centering \resizebox{1.00\hsize}{!}{\includegraphics{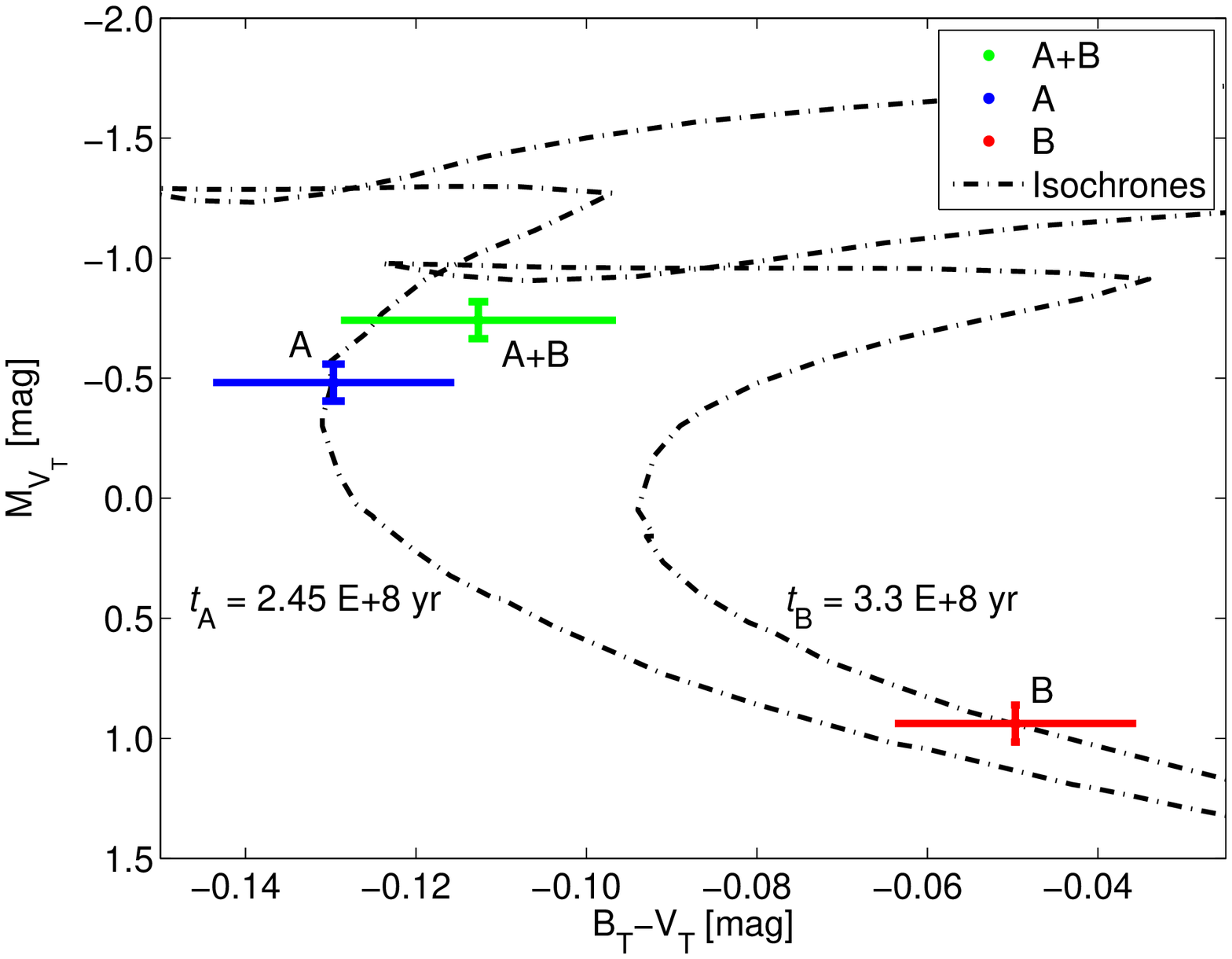}}
            \caption{Detail of HR diagram with positions of \pdra~compiled on the Tycho magnitudes from \citet{fabricius2000} and parallax from \citet{vanleeuwen2007}. 
			 }
            \label{fig-hrdiagram}
        \end{figure}

Measurements from Hipparcos satellite (parallax, and two-colour photometry from Tycho instrument) were utilized for construction of HR diagram. We adopted information about brightnesses of A and B components of \pdra~from \citet{fabricius2000}, and corrected Hipparcos value of parallax from \citet{vanleeuwen2007}. The absolute magnitudes and colour indexes were corrected for interstellar extinction $A_{\rm V} = 0.123$\,mag \citep{malkov2012} using approach described in \citet{kallrath2009}. Location of both components of \pdra~in HR diagram is based on Table~\ref{tab-hrdiagram}. They belong among high luminous stars on the MS and only small detail of HR diagram in their vicinity is displayed in Fig.~\ref{fig-hrdiagram}. Brighter A component is dominant and thus, it is close to the position, where is the object with total luminosity, as a whole system \pdra. Fainter B component is cooler and probably less evolved star. HR diagram contains synthetic PARSEC isochrones \citep{bressan2012} which were generated according to the positions of A and B components in the diagram and with the assumption of solar metallicity. Both components lie on the different isochrones with time difference only $8.5\times 10^{7}$\,yr ($t_{\rm A} = 2.45\times 10^{8}$\,yr, $t_{\rm B} = 3.3\times 10^{8}$\,yr). Nevertheless, the selection of correct isochrone is high-dependent on the uncertainty in the colour of fainter component and the recorded distance between isochrones is low (B component is distant only about $1.5\,\sigma$ from A isochrone in $B_{\rm T}-V_{\rm T}$). Probability that two angularly close stars have almost the same age is too low, thus, we consider it as the possible evidence for the gravitational bond between A and B components.

\begin{center}
\begin{table}
\begin{center}\caption{Parameters of binary pair \pdra~AB used in HR diagram.}
\begin{tabular}{lccc}
\hline
Component						& A				& B				& A+B$^{*}$\\
\hline
$B_{\rm T} - V_{\rm T}$ [mag]	& $-0.13(1)$	& $-0.05(1)$	& $-0.113(16)$\\ 
$M_{V_{\rm T}}$ [mag]			& $-0.48(8)$	& 0.94(8)		& $-0.74(8)$\\ 
Parallax [mas]	 & \multicolumn{3}{c}{10.77(38)} \\
Distance [pc]	 & \multicolumn{3}{c}{92.9(33)} \\
$A_{\rm V}$ [mag]	 & \multicolumn{3}{c}{0.123} \\
\hline
\end{tabular}\label{tab-hrdiagram}
\end{center}
{\bf Notes:} $^{(*)}$ Values for the whole system A+B were calculated from parameters for A and B components and from Pogson equation.\\
\end{table}
\end{center}

\section{Summary and conclusions}\label{summarysection}
We performed a complete analysis of the multiple stellar system \pdra~(expected hierarchy of the system is in Fig.~\ref{fig-hierarchy}) with focus on the inner binary system Aab. We disproved orbital elements from \citet{abt1973}, and utilizing 4 sources of RV measurements we found 128-day period and other orbital parameters which are close to results determined by \citet{beardsley1969}. 


The orbit of the outer visual binary system AB was solved three-times during history. We showed that orbital elements from \citet{olevic1975} and \citet{olevic1990} do not describe new measurements. The latest solution from \citet{andrade2005} fits well all available relative position measurements and relative orbit of the binary. Nevertheless, these variations could be well explained using unbound binary star motion as was shown in Sect.~\ref{VisBinUnbound}. The current difference between bound and unbound models of motion B star relative to A is still small and thus unbound explanation cannot be completely ruled out. We attempted to find proofs for the bound binary hypothesis. The similarities in proper motions and the HR diagram, where A and B components lie on the almost same isochrone, we present as the most promising evidences.

We also discussed the last possible component C, but we present clear evidence that its gravitational bond is not real (e.g. different proper motions). In each case, its influence on the whole system is almost undetectable.

Visual orbit of the outer binary AB allowed us to determine the total mass of the triple system about 7.6\,M$_{\odot}$, nevertheless with respect to the older models of visual system and with the fact that only about half of the orbit have been finished from its discovery, the value must be accepted circumspectly. We adopted mass of Aa component $M_{\rm Aa} = 3.30$\,M$_{\odot}$ from literature and we obtained minimal Ab mass $M_{\rm Ab\,min} \sim 1.36$\,M$_{\odot}$ from spectroscopic orbit. The rest of total mass remains on the B component with maximal mass $M_{\rm B\,max} \sim 2.93$\,M$_{\odot}$.

The most probable isochrones in HR diagram predict the same mass of Aa component $M_{\rm Aa} \sim 3.28$\,M$_{\odot}$ and little different values for masses of other stars $M_{\rm Ab} \sim 1.91$\,M$_{\odot}$, and $M_{\rm B} \sim 2.40$\,M$_{\odot}$, which can be possible. These masses, together with the mass function obtained from the spectroscopic orbit, allow us to estimate the inclination of the Aab system orbital plane, $i_{\rm Aab} \sim 50\,^{\circ}$.

Preliminary values of masses for individual components have been already determined. Unfortunately, authors used unreliable value of inner orbital period from \citet{abt1973}, thus their values are not actual: $M_{\rm Aa} = 3.4(3)$\,M$_{\odot}$, $M_{\rm Ab} = 1.4(2)$\,M$_{\odot}$, $M_{\rm B} = 2.7(2)$\,M$_{\odot}$ \citep{docobo2006}, or $M_{\rm Aa} = 3.20$\,M$_{\odot}$, $M_{\rm Ab\,min} = 2.05$\,M$_{\odot}$, $M_{\rm B} = 0.87$\,M$_{\odot}$ \citep{tokovinin2008}\footnote{\citet{tokovinin2008} used older value of parallax 11.28(48)\,mas \citep{esa1997}. Therefore the total mass is only 6.12\,M$_{\odot}$.}.



\section*{Acknowledgements}
This research has made use of NASA's Astrophysics Data System, the Washington Double Star Catalog maintained at the U.S. Naval Observatory, the VizieR catalogue access tool and the SIMBAD database, operated at CDS, Strasbourg, France. I thank very much my colleagues, namely E.~Paunzen, M.~Prv\'{a}k, J.~Krti\v{c}ka, M.~Zejda, Z.~Mikul\'{a}\v{s}ek, and also the anonymous referee, who helped me to improve the manuscript. This work was supported by grants GA~\v{C}R 13-10589S, MUNI/A/1110/2014 and MUNI/A/1494/2014.


\end{document}